\documentclass[showpacs,showkeys]{revtex4}
\usepackage{graphicx}
\usepackage{soul}
\usepackage{color}
\usepackage{amsmath}
\usepackage{pgf}
\newcommand{\be}{\begin{equation}}
\newcommand{\bel}{\begin{equation}\label}
\newcommand{\ee}{\end{equation}}
\def\dg{\dagger}
\newcommand{\barl}{\begin{eqnarray}\label}
\newcommand{\ear}{\end{eqnarray}}

\begin{document}
\title{
Qubit lattice coherence induced by electromagnetic pulses in superconducting metamaterials}
\author{Z. Ivi\'c$^{1,2,5}$, N. Lazarides$^{1,3,5}$, G. P. Tsironis$^{1,3,4,5}$}
\affiliation{
$^{1}$Crete Center for Quantum Complexity and Nanotechnology, Department of Physics,
      University of Crete, P. O. Box 2208, 71003 Heraklion, Greece. \\
$^{2}$University of Belgrade, "Vin{\v c}a" Institute of Nuclear Sciences,
      Laboratory for Theoretical and Condensed Matter Physics, P.O.Box 522, 11001 Belgrade,
      Serbia. \\ 
$^{3}$Institute of Electronic Structure and Laser,
      Foundation for Research and Technology--Hellas, P.O. Box 1527, 71110 Heraklion,
      Greece. \\
$^{4}$Department of Physics, School of Science and Technology, Nazarbayev University,
      53 Kabanbay Batyr Ave., Astana 010000, Kazakhstan. \\
$^{5}$National University of Science and Technology MISiS, Leninsky prosp. 4, Moscow, 
      119049, Russia
}

\date{\today}
\begin{abstract} 
Quantum bits (qubits) are at the heart of quantum information processing schemes. Currently,
solid-state qubits, and in particular the superconducting ones, seem to satisfy the
requirements for being the building blocks of viable quantum computers, since they exhibit
relatively long coherence times, extremely low dissipation, and scalability.
The possibility of achieving quantum coherence in macroscopic circuits comprising
Josephson junctions, envisioned by Legett in the 1980's, was demonstrated for the first
time in a charge qubit; since then, the exploitation of macroscopic quantum effects in
low-capacitance Josephson junction circuits allowed for the realization of several kinds
of superconducting qubits. Furthermore, coupling between qubits has been successfully
achieved that was followed by the construction of multiple-qubit logic gates and the
implementation of several algorithms. Here it is demonstrated that induced qubit lattice
coherence as well as two remarkable quantum coherent optical phenomena, i.e., self-induced
transparency and Dicke-type superradiance, may occur during light-pulse propagation in
quantum metamaterials comprising superconducting charge qubits. The generated qubit lattice
pulse forms a compound "quantum breather" that propagates in synchrony with the electromagnetic
pulse. The experimental confirmation of such effects in superconducting quantum metamaterials
may open a new pathway to potentially powerful quantum computing.
\end{abstract}
\maketitle

\section*{Introduction}
Quantum simulation, that holds promises of solving particular problems exponentially
faster than any classical computer, is a rapidly expanding field of research
\cite{Devoret2013,Paraoanu2014,Georgescu2014}.
The information in quantum computers is stored in quantum bits or qubits, which have found
several physical realizations; quantum simulators have been nowadays realized and/or proposed
that employ trapped ions \cite{Blatt2012}, ultracold quantum gases \cite{Bloch2012},
photonic systems \cite{Aspuru-Guzik2012}, quantum dots \cite{Press2008}, and superconducting
circuits \cite{Houck2012,Schmidt2013,Devoret2013}.
Solid state devices, and in particular those relying on the Josephson effect \cite{Josephson1962},
are gaining ground as preferable elementary units (qubits) of quantum simulators since they
exhibit relatively long coherence times and extremely low dissipation \cite{Wendin2007}.
Several variants of Josephson qubits that utilize either charge or flux or phase degrees
of freedom have been proposed for implementing a working quantum computer;
the recently anounced, commercially available quantum computer with more than $1000$ '
superconducting qubit CPU,
known as D-Wave 2X$^{TM}$ (the upgrade of D-Wave Two$^{TM}$ with $512$ qubits CPU),
is clearly a major advancement in this direction.
A single superconducting charge qubit (SCQ) \cite{Pashkin2009} at milikelvin
temperatures behaves effectively as an artificial two-level "atom" in which two
states, the ground and the first excited ones, are coherently superposed by Josephson
coupling. When coupled to an electromagnetic (EM) vector potential, a single SCQ does
behave, with respect to the scattering of EM waves, as an atom in space.
Indeed, a “single-atom laser” has been realized with an SCQ coupled to a transmission
line resonator (“cavity”) \cite{Astafiev2007}.
Thus, it would be anticipated that a periodic structure of SCQs demonstrates the properties
of a transparent material, at least in a particular frequency band. The idea of building
materials comprising artificial "atoms" with engineered properties, i.e., {\em metamaterials},
and in particular superconducting ones \cite{Jung2014}, is currently under active
development. {\em Superconducting quantum metamaterials} (SCQMMs) comprising a large
number of qubits could hopefully maintain quantum coherence for times long enough to reveal
new, exotic collective properties. The first SCQMM that was only recently implemented
comprises $20$ flux qubits arranged in a double chain geometry \cite{Macha2014}.
Furthermore, lasing in the microwave range has been demonstrated theoretically to be triggered
in an SCQMM initialized in an easily reachable factorized state \cite{Asai2015}.

\section*{Results}
\subsection*{Superconducting Quantum Metamaterial Model}
Consider an infinite, one-dimensional (1D) periodic SCQ array placed in a transmission
line (TL) consisting of two superconducting strips of infinite length
\cite{Rakhmanov2008,Shvetsov2013} (Figure 1a and b); each SCQ, in the form of a tiny 
superconducting island, is connected to each bank of the TL by a Josephson junction (JJ).
The control circuitry for each individual SCQ (Figure 1c), consisting of a gate voltage 
source $V_g$ coupled to it through a gate capacitor $C_g$, allows for local control of 
the SCQMM by altering independently the state of each SCQ \cite{Zagoskin2011}.  
The SCQs exploit the nonlinearity of the Josephson effect and the large charging energy
resulting from nanofabrication to create artificial mesoscopic two-level systems.
A propagating EM field in the superconducting TL gives rise to nontrivial
interactions between the SCQs, that are mediated by its photons \cite{vanLoo2013}.
Those interactions are of fundamental importance in quantum optics, quantum simulations,
and quantum information processing, as well. In what follows, it is demonstrated theoretically
that self-induced transparency \cite{McCall1967} and Dicke-type superradiance (collective
spontaneous emission) \cite{Dicke1954} occur for weak EM fields in that SCQMM structure;
the occurence of the former or the latter effect solely depends on the initial state of the
SCQ subsystem. Most importantly, self-induced transparent (SIT) or superradiant (SRD) pulses
induce quantum coherence effects in the qubit subsystem.
In superradiance (resp. self-induced transparency), the initial conditions correspond to
a state where the SCQs are all in their excited (resp. ground) state;
an extended system exhibiting SRD or SIT effects is often called a coherent
amlpifier or attenuator, respectively. These fundamental quantum coherent prosesses have
been investigated extensively in connection to one- and two-photon resonant two-level systems.
Superradiant effects have been actually observed recently in two-level systems formed by
quantum dot arrays \cite{Scheibner2007} and spin-orbit coupled Bose-Einstein condensates
\cite{Hamner2014}; the latter system features the coupling between momentum states and
the collective atomic spin which is analogous to that between the EM field and the atomic
spin in the original Dicke model. These results suggest that quantum dots and the atoms
in the Bose-Einstein condensate can radiatively interact over long distances.
The experimental confirmation of SIT and SRD in extended SCQMM structures may open a new
pathway to potentially powerful quantum computing. As a consequence of these effects,
the value of the speed of either an SIT or SRD propagating pulse in a SCQMM structure can
in principle be engineered through the SCQ parameters \cite{Cornell2001}, which
{\em is not possible in ordinary resonant media}.
From a technological viewpoint, an EM (light) pulse can be regarded as a "bit" of optical
information; its slowing down, or even its complete halting for a certain time interval,
may be used for data storage in a quantum computer.
\begin{figure}[ht]
   \centering
   \includegraphics[width=0.75\linewidth]{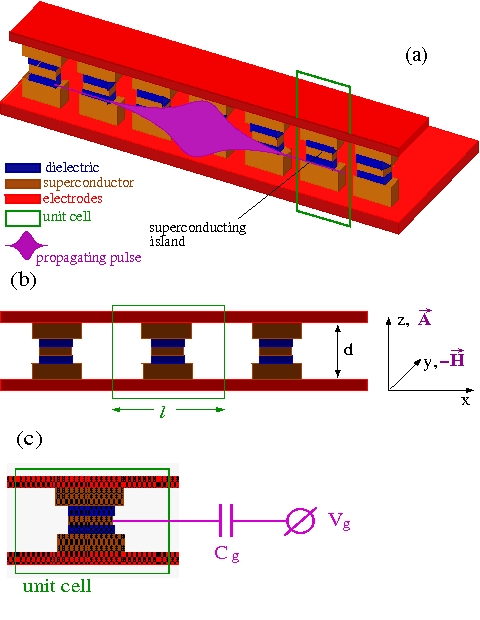}
\caption{
{\bf Schematic drawing of a charge qubit superconducting quantum metamaterial (SCQMM).}
({\bf a}) The SCQMM comprising an infinite chain of identical charge qubits in a 
superconducting transmission line. Each qubit consists of a superconducting island that 
is connected to the electrodes of the transmission line through two Josephson junctions,
formed in the regions of the dielectric layers (blue). The propagating electromagnetic
vector potential pulse is also shown schematically out of scale.
({\bf b}) The side view of the SCQMM in which the relevant geometrical parameters and the
field orientations are indicated.
({\bf c}) A unit cell of the superconducting quantum metamaterial which also shows the 
control circuitry of the charge qubit, consisting of a gate potential $V_g$ applied to it
through the gate capacitor $C_g$. 
}
\label{fig1}
\end{figure}

In the following, the essential building blocks of the SCQMM model are summarized in a
self-contained manner, yet omitting unnecessary calculational details which are presented
in the Supplementary Information.
The energy per unit cell of the SCQMM structure lying along the $x-$direction, when coupled
to an EM vector potential $\vec{A}=A_z (x,t) \hat{z}$, can be readily written as
\cite{Rakhmanov2008,Shvetsov2013}
\begin{eqnarray}
\label{1}
   H= \sum_n \left\{ \left[ \dot\varphi^2_n-2\cos \varphi_n \right]
            +\left[ \dot\alpha^2_n+\beta^2(\alpha_{n+1}-\alpha_n)^2 \right] \right.
           \nonumber \\ \left.
            +\left[ 2\cos \varphi_n (1-\cos \alpha_n) \right] \right\},
\end{eqnarray}
in units of the Josephson energy $E_J=\Phi_0 I_c/(2\pi C)$, with $\Phi_0$, $I_c$ and $C$
being the magnetic flux quantum, the critical current of the JJ, and the capacitance of
the JJ, respectively. In equation (\ref{1}), $\varphi_n$ is the superconducting phase on the $n$th
island, $\beta=(8\pi d E_J)^{-1/2} (\Phi_0/2\pi)$, with $d$ being the separation between
the electrodes of the superconducting TL,
and the overdots denote differentiation with respect to the temporal variable $t$.
Assuming EM fields with wavelengths $\lambda >> \ell, d$, with $\ell$ being the distance
between neighboring qubits, the EM potential is approximately constant within a unit cell,
so that in the centre of the $n$th unit cell $A_z (x,t) \simeq A_{z,n} (t)$.
In terms of the discretized EM potential $A_{z,n} (t)$, the normalized gauge term is
$a_n =2\pi d A_{x,n} /\Phi_0$. The classical energy expression equation (\ref{1}) provides a
minimal modelling approach for the system under consideration; the three angular
brackets in that equation correspond to the energies of the SCQ subsystem, the EM field
inside the TL electrodes, and their interaction, respectively.
The latter results from the requirement for gauge-invariance of each Josephson phase.

\subsection*{Second Quantization and Reduction to Maxwell-Bloch Equations}
The quantization of the SCQ subsystem requires the replacement of the classical variables
$\varphi_n$ and $\dot\varphi_n$ by the corresponding quantum operators $\hat{\varphi}_n$
and $-i({\partial}/{\partial \hat{\varphi}_n})$, respectively. While the EM field is treated
classically, the SCQs are regarded as two-level systems, so that only the two lowest energy
states are retained; under these considerations, the second-quantized Hamiltonian
corresponding to equation (\ref{1}) is
\begin{eqnarray}
\label{2}
   H=\sum_n \sum_{p} E_{p}(n) a^{\dagger}_{n,p} a_{n,p} 
    +\sum_n \left[ \dot\alpha^2_n+\beta^2(\alpha_{n+1}-\alpha_n)^2 \right] \nonumber \\
    +4\sum_n \sum_{p,p'} V_{p,p'}(n) a^{\dagger}_{n,p} a_{n,p'} \sin^2\frac{\alpha_n}{2},~~~~~
\end{eqnarray}
where $p,p'=0,1$, $E_0$ and $E_1$ are the energy eigenvalues of the ground and the excited
state, respectively, the operator $a^{\dagger}_{n,p}$ ($a_{n,p}$) excites (de-excites)
the $n$th SCQ from the ground to the excited (from the excited to the ground) state,
and $V_{p,p'}=\int d\varphi \Xi^*_p(\varphi) \cos \varphi \Xi_p(\varphi)$ are the matrix
elements of the effective SCQ-EM field interaction. The basis states $\Xi_{p}$ can
be obtained by solving the single-SCQ Schr\"odinger equation
$(-{\partial^2}/{\partial \varphi^2} -E_p +2\cos\varphi ) \Xi_p=0$.
In general, each SCQ is in a superposition state of the form
$|\Psi_n\rangle=\sum_p\Psi_{n,p}(t)a^{\dagger}_{n,p}|0\rangle$. The substitution of
$|\Psi_n\rangle$ into the Schr{\"o}dinger equation with the second-quantized Hamiltonian
equation (\ref{2}), and the introduction of the Bloch variables
$R_x (n)=\Psi^\star_{n,1} \Psi_{n,0} +\Psi^\star_{n,0} \Psi_{n,1}$,
$R_y (n)=i(\Psi^\star_{n,0} \Psi_{n,1} -\Psi^\star_{n,1} \Psi_{n,0})$,
$R_z (n)=|\Psi_{n,1}|^2 -|\Psi_{n,0}|^2$,
provides the re-formulation of the problem into the Maxwell-Bloch (MB) equations
\begin{eqnarray}
\label{3a}
  \dot{R}_x(n)&=-\left[ \Delta +8 D \sin^2 \frac{\alpha_n}{2} \right] R_y (n), \\
\label{3b}
  \dot{R}_y(n)&=\left[ \Delta +8 D \sin^2 \frac{\alpha_n}{2} \right] R_x (n)
                                     -8\mu \sin^2 \frac{\alpha_n}{2} R_z (n) , \\
\label{3c}
  \dot{R}_z(n)&=+8\mu \sin^2 \frac{\alpha_n}{2} R_y (n) , 
\end{eqnarray}
that are {\em nonlinearly} coupled to the resulting EM vector potential equation
\begin{equation}
\label{4}
  \ddot\alpha_n 
   +\left\{ \Omega^2 + \chi \left[ \mu R_x (n) +D R_z (n) \right] \right\} \sin \alpha_n 
   =\beta^2 \delta a_n ,
\end{equation}
where $\delta \alpha_n =\alpha_{n-1} -2\alpha_n +\alpha_{n+1}$, $D=(V_{11} -V_{00})/(2\chi)$,
$\Omega^2 =( V_{00} +V_{11} ) / 2$, $\mu=V_{10}/\chi =V_{01}/\chi$,
and $\Delta=\epsilon_1 -\epsilon_0 \equiv (E_1-E_0) / \chi$, with $\chi=\hbar \omega_J/E_J$.
In the earlier equations, the overdots denote differentiation with respect to the normalized
time $t \rightarrow \omega_J t$, in which $\omega_J = e I_c/(\hbar C)$ is the Josephson
frequency and $e$, $\hbar$ are the electron charge and the Planck's constant devided by
$2\pi$, respectively.
\begin{figure}[ht]
   \centering
   \includegraphics[width=0.75\linewidth]{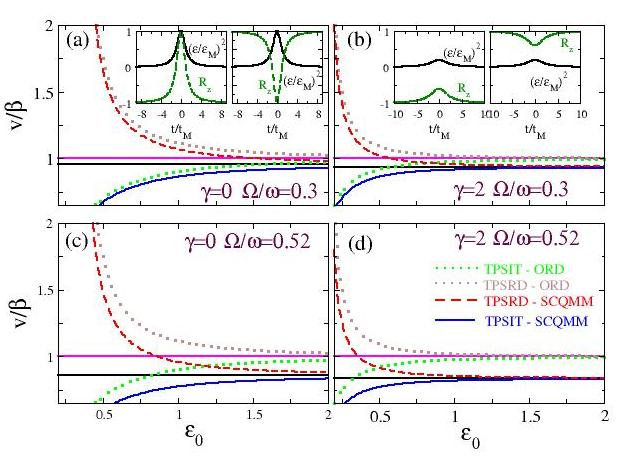}
\caption{
{\bf The velocity-amplitude relation in two-photon superradiant (TPSRD, amplifying)
and two-photon self-induced transparent (TPSIT, absorbing) superconducting quantum
metamaterials (SCQMMs) $\&$ quantum coherent pulse profiles.}
In all subfigures, the pulse velocity $v$ in units of $\beta$ as a function of the
electromagnetic vector potential pulse amplitude $\varepsilon_0$ is plotted and compared
with the corresponding curves for ordinary (atomic) amplifying and absorbing media
(brown- and green-dotted curves, respectively).
The horizontal magenta-solid (resp. black-solid) lines indicate the limiting velocity in
ordinary amplifying and absorbing media, $v/\beta=1$ (resp. amplifying and absorbing
SCQMMs, $v =c < \beta$).
({\bf a}) $V_{00} =V_{11} =1$, $V_{01} =V_{10} =0.8$, $\chi=1/5$, $E_1 -E_0=3$
   ($\gamma=0$ and $\Omega/\omega=0.3$).
   Left Inset: The electromagnetic vector potential pulse envelop
   $(\varepsilon/\varepsilon_M)^2$ and the population inversion function $R_z (n)$
   profiles as a function of the slow variable $(\tau/\tau_M$ in a frame of reference that
   is moving with velocity $v$, for TPSIT (absorbing) SCQMMs.
   Right Inset: Same as in the left inset for TPSRD (amplifying) SCQMMs.
({\bf b}) $V_{00}=0.6$, $V_{11} =1.4$, $V_{01} =V_{10} =0.8$, $\chi=1/5$, $E_1 -E_0=3$
   ($\gamma=2$ and $\Omega/\omega=0.3$).
   Left Inset: The electromagnetic vector potential pulse envelop
   $(\varepsilon/\varepsilon_M)^2$ and the population inversion function $R_z (n)$
   profiles as a function of the slow variable $(\tau/\tau_M$ in a frame of reference that
   is moving with velocity $v$, for TPSIT (absorbing) SCQMMs in the presence of relatively
   strong decoherence ($\gamma=2$).
   Right Inset: Same as in the left inset for TPSRD (amplifying) SCQMMs.
({\bf c}) $V_{00} =V_{11} =3$, $V_{01} =V_{10} =0.8$, $\chi=1/5$, $E_1 -E_0=3$
   ($\gamma=0$ and $\Omega/\omega=0.52$).
({\bf d}) $V_{00} =3$, $V_{11} =3.8$, $V_{01} =V_{10} =0.8$, $\chi=1/5$, $E_1 -E_0=3$
   ($\gamma=2$ and $\Omega/\omega=0.52$).
The effect of non-zero decoherence ($\gamma \neq 0$) become apparent by direct comparison
of {\bf a} with {\bf b} and {\bf c} with {\bf d}.
The pulse velocity $v$ in SCQMMs saturates with increasing $\varepsilon_0$ to $v_m/\beta$,
that can be significantly lower than that achieved in ordinary TPSIT and TPSRD media, i.e.,
$\beta$.
The parameters of the SCQMM can be engineered to slow down the pulse velocity $v$ at the
desired level for high enough amplitudes $\varepsilon_0$. Note that $v$ is also the velocity
of the coherent qubit pulse.
}
   \label{fig2}
\end{figure}

\subsection*{Approximations and Analytical Solutions}
For weak EM fields, the approximation $\sin\alpha_n \simeq \alpha_n$ can be safely used.
Then, by taking the continuum limit $\alpha_n (t) \rightarrow \alpha(x,t)$ and
$R_i (n; t) \rightarrow R_i (x; t)$ $(i=x,y,z)$ of equations (\ref{3a}-\ref{3c}) and (\ref{4}),
a set of simplified, yet still nonlinearly coupled equations is obtained, similar to those
encountered in {\em two-photon} SIT in resonant media \cite{Belenov1969}.
Further simplification can be achieved with the slowly varying envelope approximation (SVEA)
by making for the EM vector potential the ansatz $\alpha(x,t)=\varepsilon(x,t)\cos \Psi(x,t)$,
where $\Psi(x,t)=k x-\omega t+\phi(x,t)$ and $\varepsilon(x,t)$, $\phi(x,t)$ are the
slowly varying pulse envelope and phase, respectively, with $\omega$ and
$k=\pm \sqrt{\omega^2-\Omega^2}/\beta$ being the frequency of the carrier wave of the EM
pulse and its wavenumber in the superconducting TL, respectively. In the absence of the
SCQ chain the EM pulse is "free" to propagate in the TL with speed $\beta$.
At the same time, equations (\ref{3a}-\ref{3c}) for the Bloch vector components are transformed
according to
$R_x = r_x \cos (2\Psi) +r_y \sin(2\Psi)$, $R_y= r_y\cos(2\Psi) -r_x \sin(2\Psi)$, and
$R_z =r_z$. Then, collecting the coefficients of $\sin\Psi$ and $\cos\Psi$ while neglecting
the rapidly varying terms, and averaging over the phase $\Psi$, results in a set of
truncated equations (see Supplementary Information). Further manipulation of the resulting
equations and the enforcement of the {\em two-photon resonance condition} $\Delta=2\omega$,
results in
\begin{eqnarray}
\label{5.1}
  \dot{\varepsilon} +c \varepsilon_x =-\chi \frac{\mu}{\Delta} \varepsilon r_y , \\
\label{5.2}
  \dot{\phi} +c \phi_x =-\chi \frac{2 D}{\Delta} r_z ,
\end{eqnarray}
where $c=\beta^2 k/\omega =2 \beta^2 k/\Delta$, and the truncated MB equations
\begin{eqnarray}
\label{5}
   \dot{r}_x=-2 D \varepsilon^2 r_y, 
   \dot{r}_y=+2 D \varepsilon^2 r_x -\frac{\mu\varepsilon^2}{2} R_z,
   \dot{r}_z=+\frac{\mu\varepsilon^2}{2} r_y,
\end{eqnarray}
which obey the conservation law $r_x^2 +r_y^2 +r_z^2 =1$. In equations (\ref{5}), the
$n-$dependence of the $r_i$ $(i=x,y,z)$ is suppressed, in accordance with common
practices in quantum optics.

The $r_i$ can be written in terms of new Bloch vector components $S_i$ using the unitary
transformation $r_x =S_x \cos\Phi -S_z \sin\Phi$, $r_y =S_y$, and
$r_z =S_z \cos\Phi +S_x \sin\Phi$, where $\Phi$ is a constant angle to be determined.
Using a procedure similar to that for obtaining the $r_i$, we get
$\dot S_x =0$, $\dot S_y =-\frac{1}{2} W \varepsilon^2 S_z$, and
$\dot S_z =+\frac{1}{2} W \varepsilon^2 S_y$, where $W=\sqrt{(4D)^2 +\mu^2}$ and
$\tan \Phi \equiv \gamma ={4 D}/{\mu}$.
The combined system of the equations for the $S_i$ and equations (\ref{5.1})-(\ref{5.2}) admits
exact solutions of the form $\varepsilon=\varepsilon(\tau=t-x/v)$ and $S_i =S_i (\tau=t-x/v)$,
where $v$ is the pulse speed. For the slowly varying pulse envelop, we obtain
\begin{eqnarray}
\label{6}
  \varepsilon (\tau) =\varepsilon_0 
     \left[ 1 +\left( \frac{\tau -\tau_0}{\tau_p} \right)^2 \right]^{-\frac{1}{2}} ,    
\end{eqnarray}
where $\varepsilon_0 =\sqrt{ (4\sigma^2/\omega) [ v / (c-v) ] }$ is the pulse amplitude and
$\tau_p =\left\{ \chi (\sigma \mu/\omega) [ v / (c-v) ] \right\}^{-1}$ its duration,
with $\sigma =\mu/W =1/\sqrt{1+\gamma^2}$. The decoherence factor $\gamma$ can be expressed
as a function of the matrix elements of the SCQ-EM field interaction, $V_{ij}$, as
$\gamma=2 (V_{11} -V_{00})/V_{10}$ that can be calculated when the latter are known.
Such Lorentzian propagating pulses have been obtained before in two-photon resonant media
\cite{Tan-no1975a,Nayfeh1978}; however, SIT in quantum systems has only been
demonstrated in one-photon (absorbing) frequency gap media, in which solitonic pulses can
propagate without dissipation \cite{John1999}.
The corresponding solution for the population inversion, $R_z$, reads
\begin{equation}
\label{7}
  R_z (\tau) =\pm\left[-1 +\left( \frac{\varepsilon (\tau)}{\varepsilon_M} \right)^2
                 \right] ,
\end{equation}
where $\varepsilon_M =2\sqrt{ (1/\omega) [ v / (c-v) ] }$, and the plus (minus) sign
corresponds to absorbing (amplifying) SCQMMs; these are specified through the initial
conditions as
$R_z(-\infty)=-1$, $\varepsilon(-\infty)=0$ and $R_z(-\infty=+1)$, $\varepsilon(-\infty)=0$
for absorbing and amplifying SCQMMs, respectively
(with $R_x (-\infty) =R_y (-\infty)=0$ in both cases).
The requirement for the wavenumber $k$ being real, leads to the SCQ parameter-dependent
condition $2 \chi^2 (V_{11}+V_{00}) < (E_1-E_0)^2$ for pulse propagation in the SCQMM.
Thus, beyond the obtained two-photon SIT or SRD, the propagating EM pulse plays
a key role in the interaction processes in the qubit subsystem: it leads to collective
behavior of the ensemble of SCQs in the form of quantum coherent probability pulses;
such pulses are illustrated here through the population inversion $R_z$.

The corresponding velocity-amplitude relation of the propagating pulse reads
\begin{eqnarray}
\label{8}
   v= c \left[1 \pm { \chi \frac{4 \sigma^2}{\omega \varepsilon_0^2}} \right]^{-1} .
\end{eqnarray}
Equation (\ref{8}) can be also written as a velocity-duration expression, since the pulse
amplitude and its duration are related through $\varepsilon_0^2 \tau_p =4/(\chi W)$.
The duration of SRD pulses cannot exceed the limiting value of $\tau_M=\omega (c-v)/(\chi \mu v)$.
From equation (\ref{8}), the existence of a critical velocity $c$, defined earlier,
can be immediately identified; that velocity sets an upper (lower) bound on the pulse
velocity in absorbing (amplifying) SCQMM structures. Thus, in absorbing (amplifying) SCQMM
structures,
pulses of higher intensity propagate faster (slower). That limiting velocity is generally
lower than the corresponding one for two-photon SIT or SRD in ordinary media, $\beta$,
which here coincides with the speed of the "free" pulse in the TL (Figure 2).
As can be inferred from Figure 2, the increase of decoherence through $\gamma$ makes
the velocity to saturate at its limiting value $c$ at lower amplitudes $\varepsilon$;
that velocity can be reduced further with increasing the ratio of the TL to the
pulse carrier wave frequency $\Omega/\omega$ through proper parameter engineering.
Moreover, effective control of $v$ in SCQMMs could in principle be achieved by an external
field \cite{Park2001} or by real time tuning of the qubit parameters.
That ability to control the flow of "optical", in the broad sense, information may have
technological relevance to quantum computing \cite{Cornell2001}.
Note that total inversion, i.e. excitation or de-excitation of all qubits during pulse
propagation is possible only if $\gamma=0$, i.e., for $V_{00}=V_{11}$; otherwise
($V_{00} <V_{11}$) the energy levels of the qubit states are Stark-shifted, violating thus
the resonance condition.
Typical analytical profiles for the EM vector potential pulse $\varepsilon(\tau)$
and the population inversions $R_z (\tau)$ both for absorbing and amplifying SCQMMs are
shown in the insets of Figure 2. The maximum of $\varepsilon(\tau)$ reduces considerably
with increasing $\gamma$, while at the same time the maximum (minimum) of $R_z$ decreases
(increases) at the same rate.

The system of equations (\ref{5.1})-(\ref{5.2}) and (\ref{5}) can be reduced to a single 
equation using the parametrization 
$r_x =R_0 \gamma \sigma^2 [1 -\cos\theta]$, $r_y =-R_0 \sigma \sin\theta$, and 
$r_z =R_0 \{ 1 -\sigma^2 [1 -\cos\theta] \}$, of the Bloch vector components.
Then, a relation between the Bloch angle $\theta \equiv \theta(x,t)$ and the slow amplitude
$\varepsilon$ can be easily obtained, that leads straightforwardly to the equation
$\ddot\theta +c\dot{\theta}_x =-R_0 \chi \frac{\mu}{\omega} \frac{\partial}{\partial t}
\cos\theta$. Time integration of that equation yields $\frac{\partial \theta}{\partial x}
=R_0 \chi \frac{\mu}{c \omega} (1 -\cos\theta)$, that conforms with the famous {\em area theorem:}
{\em pulses with special values of "area" $\theta (x)=2\pi n$ conserve that value during
propagation}.

Here we concentrate on the interaction of the SCQs with the EM wave and we are not concerned
with decoherence effects in the SCQs due to dephasing and energy relaxation. This is clearly 
an idealization which is justified as long as the coherence time exceeds the wave propagation 
time across a relatively large  number of unit cell periods. In a recent experiment 
\cite{Gambetta2006}, a charge qubit coupled to a strip line had a dephasing time in excess 
of $200 ~ns$, i.e., a dephasing rate of $5~MHz$, and a photon loss rate from the cavity of 
$0.57 ~MHz$. Those frequencies are very small compared with the transition frequency of the 
considered SCQs which is of the order of the Josephson energy (i.e., a few $GHz$) 
\cite{Rakhmanov2008,Shvetsov2013}. Therefore, we have neglected such decoherence effects 
in the present work. The  decoherence factor $\gamma$, which in Figures 2b and 2d has been 
chosen according to the parameter values in \cite{Rakhmanov2008}, is not related to either
dephasing or energy relaxation. That factor attains a non-zero value whenever the matrix
elements of the effective SCQ-EM field interaction, $V_{11}$ and $V_{00}$, are not equal.

\begin{figure}[ht]
   \centering
   \includegraphics[width=0.95\linewidth]{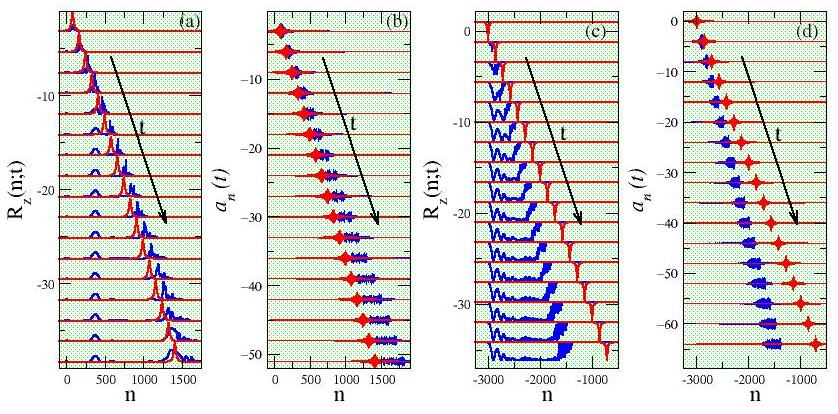}
\caption{
{\bf Numerical validation of the analytical expressions for two-photon self-induced
transparent (TPSIT) and superradiant (TPSRD) propagating pulses.}
({\bf a}) Snapshots of the population inversion pulse $R_z (n;t)$, excited by the induced
quantum coherence in the qubit subsystem by the electromagnetic vector potential pulse,
in the absence of decoherence ($\gamma=0$); the pulse propagates to the right
(time increases downwards) in TPSIT (absorbing) superconducting quantum metamaterials
(SCQMMs).
The snapshots are taken at intervals of $20$ time-units starting at $t=20$ and they are
displaced vertically to avoid overlapping (blue pulses).
The corresponding pulses from the analytical expression equation (\ref{7}) at the same
time-instants are shown in red.
({\bf b}) Snapshots for the corresponding evolution of the electromagnetic vector potential
pulse $a_n (t)$, that exhibits significant broadening as time passes by;
the numerical and analytical pulses are shown in blue and red color, respectively.
({\bf c}) The same as in {\bf a} in TPSRD (amplifying) superconducting quantum metamaterials.
The resulting propagation is not as simple as expected from the theoretical analysis;
instead of a population inversion pulse, it is observed a rather kink-like front propagating
to the the right (blue) with a velocity considerably less than that predicted analytically for
the pulse, which analytical form is shown in red.
({\bf d}) The same as in {\bf b} in TPSRD (amplifying) superconducting quantum metamaterials.
The velocity of the $a_n (t)$ pulse (blue) is the same as that of the propagating population
inversion front, $R_z (n;t)$; however, it exhibits less broadening with time in comparison
with the corresponding numerical $a_n (t)$ pulse in {\bf b}.
The predicted analytical form is shown in red.
Parameter values: $\chi=1/5$, $\beta=6$, $V_{00}=V_{11}=1$, $V_{01}=V_{10}=0.8$, $E_1 -E_0=3$,
and $v/c=0.7$ (for {\bf a} and {\bf b}); $v/c=1.25$ (for {\bf c} and {\bf d}).
}
   \label{fig3}
\end{figure}

\subsection*{Numerical Simulations}
In order to confirm numerically the obtained results, the equations (\ref{3a})-(\ref{3c}) 
and (\ref{4}) are integrated in time using a fourth order Runge-Kutta algorithm with constant
time-step. For pulse propagation in absorbing SCQMMs, all the qubits are initially set
to their ground state while the vector potential pulse assumes its analytical form for the
given set of parameters. A very fine time-step and very large qubit arrays are used to
diminish the energy and/or probability loss and the effects of the boundaries during
propagation, respectively.
The subsequent temporal evolution in two-photon SIT SCQMM, as can be seen in Figures
3a and 3b, in which
several snapshots of the population inversion $R_z (n;t)$ and the vector potential pulses
$a_n (t)$, respectively, are shown, reveals that the latter are indeed capable of inducing
quantum coherent effects in the qubit subsystem in the form of population inversion pulses!
In Figure 3a, the amplitude of the $R_z (n;t)$ pulse gradually grow to the expected maximum
around unity in approximately $60$ time units, and they continue its course almost coherently 
(although with fluctuating amplitude) for about $160$ more time units, during which they move 
at the same speed as the vector potential pulse (Figure 3b). However, due to the inherent 
discreteness in the qubit subsystem and the lack of inter-qubit coupling, the $R_z (n;t)$ 
pulse splits at certain instants leaving behind small "probability bumps" that get pinned 
at particular qubits. After the end of the almost coherent propagation regime, the $R_z (n;t)$ 
pulse broadens and slows-down until it stops completely. At the same time, the width of the 
$a_n (t)$ pulse increases in the course of time due to discreteness-induced dispersion.
A comparison with the corresponding analytical expressions reveals fair agreement during
the almost coherent propagation regime, although both the $R_z (n;t)$ and
$a_n (t)$ pulses travel slightly faster than expected from the analytical predictions.
The temporal variable here is normalized to the inverse of the Josephson frequency 
$\omega_J$ which for typical parameter values is of the order of a few $GHz$ 
\cite{Rakhmanov2008}. Then, the almost coherent induced pulse regime in the particular 
case shown in Figure 3 lasts for $\sim 160 \times 10^{-9}$ s, or $\sim 160~ns$, which is
of the same order as the reported decoherence time for a charge qubit in \cite{Gambetta2006}
(i.e., $200~ns$).

The situation seems to be different, however, in the case of two-photon SRD pulses, as can
be observed in the snapshots shown in Figures 3c and 3d for $R_z (n;t)$ and $a_n (t)$,
respectively. Here, the lack of the
inter-qubit interaction is crucial, since the SCQs that make a transition from the excited
to their ground state as the peak of the $a_n (t)$ pulse passes by their location, cannot
return to their excited states after the $a_n (t)$ pulse has gone away. It seems, thus, that
the $a_n (t)$ pulse creates a type of a kink-like front that propagates at the same velocity.
It should be noted that the common velocity of the $R_z (n;t)$ and $a_n (t)$ pulses is
considerably different (i.e., smaller) than the analytically predicted one, as it can be
inferred by
inspection of Figures 3c and 3d. Even more complicated behavioral patterns of two-photon
SRD propagating pulses and the effect of non-zero decoherence factor are discussed in the
Supplementary Information.

\section*{Conclusion}
An SCQMM comprising SCQs loaded periodically on a superconducting TL has been investigated
theoretically using a minimalistic one-dimensional model following a semiclassical approach. 
While the SCQs are regarded as two-level quantum systems, the EM field is treated classically.
Through analytical techniques it is demonstrated that the system allows self-induced 
transparent and superradiant pulse propagation given that a particular constraint is 
fulfilled. Most importanty, it is demonstrated that the propagating EM pulses may induce
quantum coherent population inversion pulses in the SCQMM. Numerical simulation of the 
semiclassical equations confirms the excitation of population inversion pulses with 
significant coherence time in absorbing media. The situation is slightly different 
in amplifying media, in which the numerically obtained, induced population inversion 
excitations are kink-like propagating structures (although more complex behaviors discussed 
in the Supplementary Information also appear).  
Moreover, the limiting pulse velocity in both amplifying and absorbing SCQMMs is lower
than the corresponding one in two-photon resonant amplifying and absorbing ordinanary
(atomic) media. That limiting velocity in SCQMMs can in principle be engineered through
the SCQ parameters.


\section*{Acknowledgements}
This work was partially supported by
the European Union Seventh Framework Programme (FP7-REGPOT-2012-2013-1) under grant
agreement n$^o$ 316165,
the Serbian Ministry of Education and Science under Grants No. III--45010, No. OI--171009,
the Ministry of Education and Science of the Republic of Kazakhstan (Contract No. 339/76-2015),
and
the Ministry of Education and Science of the Russian Federation in the framework of the
Increase Competitiveness Program of NUST  "MISiS"(No. K2-2015-007).
\section*{Author Contributions}
Z.I., N.L., and G.P.T. performed the research, analyzed the results and wrote the paper.
\section*{Additional information}
{\bf Competing financial interests:} The authors declare no competing financial interests.


\newpage

\centerline{\bf SUPPLEMENTARY INFORMATION:}
\centerline{\bf QUBIT LATTICE COHERENCE INDUCED BY ELECTROMAGNETIC PULSES}
\centerline{\bf IN SUPERCONDUCTING METAMATERIALS} 

\section{Hamiltonian function and Quantization of the Qubit Subsystem}
Consider an infinite number of superconducting charge qubits in a 
transmission line (TL) that consists of two superconducting plates separated by 
distance $d$, while the center-to-center distance between qubits, $\ell$, is of 
the same order of magnitude. The charge qubit is of the form of a mesoscopic 
superconducting island which is connected to each electrode of the TL with a 
Josephson junction (JJ).
Assume that an electromagnetic (EM) wave corresponding to a vector potential 
$\vec{A}=A_z (x,t) \hat{z}$
propagates along the superconducting TL, in a direction parallel to 
the superconducting electrodes and perpendicular to the direction of the EM wave 
propagation. A minimalistic description of that system constists of writing the
Hamiltonian of the compound qubit array - EM field system with the geometry shown
in Fig. 1 of the paper, as
\bel{3331}
H=\sum_n \left\{ \dot\varphi^2_n-2\cos \alpha_n\cos\varphi_n+\dot\alpha^2_n 
                +\beta^2(\alpha_{n+1}-\alpha_n)^2 \right\} ,
\ee
where $\varphi_n$ is the superconducting phase on $n-$th island, 
$\beta^2=(8\pi d E_J)^{-1}  \left( \Phi_0/(2\pi) \right)^2$, with $d$ 
being the separation between the superconducting electrodes of the TL, and the 
overdots denote derivation with respect to the temporal variable $t$.
In what follows, a basic assumption is that the wavelength $\lambda$ of the EM
field is much larger than the other length scales determined by the size of the
unit cell of the qubit metamaterial ($\lambda >> \ell, d$), so that the vector
potential component $A_z (x,t)$ can be regarded to be approximatelly constant 
within a unit cell, $A_z (x,t) \simeq A_{z,n} (t)$. Then, the discretized and
normalized EM vector potential at the $n-$th unit cell which appears in Eq. 
(\ref{3331}) reads $a_n (t) =(2\pi d/\Phi_0) A_{n} (t)$. The Hamiltonian 
function Eq. (\ref{3331}) is given in units of the Josephson energy
$E_J =(\Phi_0 I_c)/(2\pi C)$, where $I_c$ and $C$ is the critical current 
and capacitance, respectively, of the JJs, and $\Phi_0 =h/(2 e)$ is the 
flux quantum, with $h$ and $e$ being the Planck's constant and the electron charge, 
respectively. That Hamiltonian can be written in a more transparent form by adding 
and subtracting $2\cos \phi_n$ and subsequently rearranging to get
\bel{3332}
   H=H_{QUB}+H_{EMF} +H_{int} ,
\ee
where the qubit subsystem energy $H_{QUB}$, the EM field energy $H_{EMF}$,
and their interaction energy $H_{int}$, take respectively the form
\bel{3}
        H_{QUB}=\sum_n \{ \dot\varphi^2_n-2\cos \varphi_n\} , 
 \qquad H_{EMF}=\sum_n \{ \dot\alpha^2_n+\beta^2(\alpha_{n+1}-\alpha_n)^2\} ,
 \qquad H_{int}=\sum_n \{ 2\cos \varphi_n (1-\cos \alpha_n)\} .
\ee
In the following, the EM field is treated classically, while the qubits are 
regarded as two-level systems. The latter approximation is particularly well
suited in the case of {\em resonant qubit - EM field interaction} adopted here.

The quantization of the qubit subsystem can be formally performed by replacing
the classical variables $\varphi_n$ and $\dot{\varphi}_n$ by the quantum 
operators $\hat{\varphi_n}$ and 
$\dot{\varphi}_n\rightarrow-i\frac{\partial}{\partial \varphi_n}$, respectively,
in the Hamiltonian $H_{QUB}$ since we are dealing with a large number of Cooper 
pairs. The exact energy spectrum $E_p (n)$ and the corresponding wavefunctions
$\Xi_p (n)$ of the $n$th qubit may then be obtained by mapping the Schr{\"o}dinger
equation with the single-qubit Hamiltonian $H_{sq}=\dot\varphi_n^2-2\cos \varphi_n$,
onto the Mathieu equation
\bel{10} 
  \bigg(\frac{\partial^2}{\partial \varphi^2_n}+E_{p,n}-2\cos\varphi_n\bigg)\Xi_{p,n}=0 .
\ee 
The second quantization of the qubit subsystem proceeds by rewritting the 
single-qubit Hamiltonian as  
\bel{99}
  H_{sq} \rightarrow H_{sq}=-\int d\varphi_n \hat{\Psi}^{\dg}(\varphi)
            \bigg(\frac{\partial^2}{\partial \varphi^2}+2\cos\varphi\bigg) 
\hat{\Psi}(\varphi) ,
\ee
where $\hat \Psi^{\dg}$ and  $\hat \Psi$ are field operators. Note that the subscript
$"n"$ in Eq. (\ref{99}) has been dropped since the qubits are identical. 
Using the expansion
$ \hat\Psi(\varphi)=\sum_p a_p \Xi_p(\varphi)$, where the operators $a^{\dg}_p$
($a_p$) create (annihilate) qubit excitations of energy $E_p$, the Hamiltonian 
Eq. (\ref{99}) is transformed into 
\barl{7'}
   H_{sq}=\sum_{p=0,1,...} E_{p} a^{\dg}_{p} a_{p} .
\ear
We hereafter restrict $H_{sq}$ to the Hilbert subspace of its two lowest levels, 
i.e., those with $p=0, 1$, so that in second quantized form the Hamiltonian
Eq. (\ref{3332}) reads
\barl{3337} 
   H= \sum_n \sum_{p} E_{p}(n)a^{\dg}_{n,p}a_{n,p} 
     +\sum_{p,p'} V_{p,p'}(n)a^{\dg}_{n,p} a_{n,p'}\sin^2\frac{\alpha_n}{2}
     +\sum_n \{ \dot\alpha^2_n+\beta^2(\alpha_{n+1}-\alpha_n)^2\} ,
 \ear
where $p, p'=0,1$ and the $V_{p,p'}(n) \equiv V_{p',p}(n)$ that represent the 
matrix elements of the $n$th qubit - EM field interaction, are given by
\bel{3338}  
   V_{p,p'}(n)=\int d\varphi_n \Xi^*_p(\varphi_n)\cos \varphi_n\Xi_{p,n}(\varphi_n) .
\ee
In the reduced state space, in which a single qubit can be either in the ground 
($p=0$) or in the excited ($p=1$) state, the normalization condition 
$\sum_p a^{\dg}_{n,p} a_{n,p}=1$ holds for each qubit in the metamaterial.
The reduced Hamiltonian Eq. (\ref{3337}) is also written in units of $E_J$.

\section{Maxwell-Bloch Formulation of the Dynamic Equations} 
In accordance with the adopted semiclassical approach, the time-dependent 
Schr\"odinger equation 
\bel{98}
   i\hbar\frac{\partial}{\partial t}|\Psi\rangle =\bar{H} |\Psi\rangle ,
\ee
in which $\bar{H}$ is the Hamiltonian from Eq. (\ref{3337}) in physical units, i.e.,
$\bar{H} =H E_J$, is employed for the description of the qubit subsystem. 
Generally, the state of each qubit is a superposition of the form
\bel{11} 
   |\Psi_n\rangle=\sum_p\Psi_{n,p}(t)a^{\dg}_{n,p}|0\rangle ,
\ee
for which it can be easily shown that the coefficients $\Psi_{n,p}$  satisfy the 
following normalization conditions
\bel{12} 
   \sum_p|\Psi_{n,p} (t)|^2 =1 , \qquad \sum_{n,p} |\Psi_{n,p}(t)|^2=N ,
\ee
where in the second Eqs. (\ref{12}) a finite $N-$qubit subsystem has been implied.
Assuming that the pulse power is not very strong, substantial simplification of 
the dynamic equations for the qubit subsystem can be achieved using the approximation
$[1-\cos(\alpha_n)] \simeq (1/2) \alpha^2_n$ in the interaction Hamiltonian 
$H_{int}$. Substitution of $|\Psi\rangle = |\Psi_n\rangle$ from Eq. (\ref{11})
into the Schr\"odinger equation (\ref{98}), along with derivation of the classical 
Hamilton's equation for the normalized EM vector potential $\alpha_n$, yields
\begin{eqnarray} 
\label{14}
&& i\dot \Psi_{n,p}=\epsilon_p\Psi_{n,p}+\frac{1}{\chi}
                     \sum_{p'}V_{p,p'}(n){\Psi}_{n,p'}\alpha^2_n, \\
\label{14.2}
&&\ddot\alpha_n-\beta^2(\alpha_{n+1}+\alpha_{n-1}-2\alpha_n)
                   +\sum_{p,p'}V_{p,p'}\Psi^{*}_{n,p}\Psi_{n,p'} \alpha_n=0 ,
\end{eqnarray}
where $\chi= \hbar\omega_j/E_J$. In Eqs. (\ref{14}) and (\ref{14.2}),
the temporal variable is renormalized according to $t \rightarrow \omega_J t$. 
Because of that, the dimensionless energy of the qubit excitations has to be 
redefined according to $E_P\rightarrow \epsilon_p=E_p/\chi$, which 
explains the presence of the dimensionless factor $1/\chi$ in Eq. (\ref{14}).

The evolution Eqs. (\ref{14}) and (\ref{14.2}) can be rewritten in terms of the 
$n-$dependent Bloch vector components through the transformation
\barl{15}
   R_z(n)=|\Psi_{n,1}|^2-|\Psi_{n,0}|^2 , \qquad
   R_y(n)=i(\Psi^*_{n,0}\Psi_{n,1}-\Psi^*_{n,1}\Psi_{n,0}) , \qquad
   R_x(n)=\Psi^*_{n,1}\Psi_{n,0}+\Psi^*_{n,0}\Psi_{n,1} . 
\ear
in which the variables $R_i$ ($i=x,y,z$) apply to each single qubit. Thus, from 
the point of view of a macroscopic system, $|\Psi_{i}|^2$ represent the
population densities
$\varrho_{i}= \frac{N}{N}|\Psi|^2\equiv \frac{N_i}{N}$. By transforming Eqs. 
(\ref{14}) and (\ref{14.2}) according to Eq. (\ref{15}) we get 
\begin{eqnarray}
\label{19.2}
   \dot R_x (n) =-(\Delta +2 D \alpha_n^2) R_y (n), \qquad
   \dot R_y (n) =+(\Delta +2 D \alpha_n^2) R_x (n) -2\mu \alpha_n^2 R_z (n), \qquad
   \dot R_z (n) =+2 \mu \alpha_n^2 R_y (n), \\
\label{20.2}
   \ddot\alpha_n + \chi [ \Omega^2 +\mu R_x (n)+D R_z (n) ] \alpha_n
                =\beta^2 ( \alpha_{n-1}-2\alpha_n+\alpha_{n+1} ) ,
\end{eqnarray}
where
\barl{18}
   D=\frac{(V_{11}-V_{00})}{2\chi} , \qquad
   \Omega^2 =\frac{(V_{11}+V_{00})}{2} , \qquad
   \mu =\frac{V_{10}}{\chi} , \qquad
   \Delta=\epsilon_1-\epsilon_0\equiv\frac{(E_1-E_0)}{\chi} .
\ear
By taking the continuous limit of Eqs. (\ref{19.2}) and (\ref{20.2}) using the 
relations $\alpha_n\rightarrow \alpha(x,t)$ and
$\alpha_{n\pm 1}\approx \alpha\pm \alpha_x+\frac{1}{2}\alpha_{xx}$, we obtain
\begin{eqnarray}
\label{19}
   \dot R_x=-(\Delta +2 D \alpha^2) R_y , \qquad 
   \dot R_y=+(\Delta +2 D \alpha^2) R_x -2\mu \alpha^2 R_z , \qquad
   \dot R_z=+2\mu\alpha^2 R_y , \\ 
\label{20}
   \ddot\alpha-\beta^2 \alpha_{xx}+\Omega^2\alpha
                           =-\chi (D R_z +\mu R_x) \alpha ,
\end{eqnarray}
where $R_x$, $R_y$, $R_z$, and $\alpha$ are functions of the spatial variable $x$
and normalized temporal variable $t$, while the overdots denote partial derivation with 
respect to the latter.
It can be easily verified that the Bloch equations Eqs. (\ref{19}) possess
the dynamic invariant $\sum_i  R^2_i =1$, whose constant value is derived through
the normalization conditions.

\section{Slowly Varying Envelope Approximation (SVEA)}
The Slowly Varying Envelope Approximation (SVEA) relies on the assumption that
the envelop of a travelling pulse in a nonlinear medium varies slowly in both time
and space compared with the period of the carrier wave, that makes possible the
introduction of slow and fast variables. Assuming that in the following,
we can approximate the EM vector potential function by
\bel{21} 
   \alpha(x,t)=\varepsilon(x,t) \cos \psi(x,t),
\ee
where $\psi(x,t)=k x-\omega t +\phi(x,t)$, with $k$ and $\omega$ being the wavenumber
and frequency of the carrier wave, respectively, that are connected through the 
dispersion relation that is obtained at a later stage. The functions $\varepsilon(x,t)$ 
and 
$\phi(x,t)$ in Eq. (\ref{21}) are the slowly varying envelop and phase, respectively.
Using fast and slow variables, the $x$ and $y$ Bloch vector components, $R_x (n)$ 
and $R_y (n)$, can be expressed as a function of new, in-phase and out-of-phase 
Bloch vector components $r_x$ and $r_y$ as
\barl{22}
  R_x=r_x \cos (2\psi) +r_y \sin(2\psi) , \qquad R_y= r_y \cos(2\psi) -r_x \sin(2\psi) ,
  \qquad R_z=r_z .
\ear

\subsection{Derivation of the dynamic equations for the slowly varying envelop and phase
of the electromagnetic vector potential}
From Eqs. (\ref{21}) and (\ref{22}), the second temporal and spatial derivative of 
$\alpha(x,t)$ can be approximated by
\barl{svea}
   \ddot{\alpha}
   \approx 2\omega \dot\varepsilon \sin\psi 
       +(2\omega \dot\phi -\omega^2) \varepsilon \cos\psi ,
   \qquad
    \alpha_{xx} \approx -2k \varepsilon_x \sin\psi-(2k \phi_x -k^2)\varepsilon \cos\psi ,
\ear
in which the rapidly varying terms of the form $\ddot\varepsilon$, $\varepsilon_{xx}$, 
$\phi^2$, $\phi_{xx}$, $\ddot\phi$, $\phi_x\varepsilon_x$, etc., have been neglected.
Substitution of Eqs. (\ref{svea}) and (\ref{22}), into Eq. (\ref{20}) gives 
\barl{100}
   2(\omega \dot\varepsilon +k\beta^2 \varepsilon_x) \sin\psi 
    +[ 2(\dot\phi \omega +k \phi_x) -\omega^2 +\Omega^2 +\beta^2 k^2 ]
         \varepsilon \cos\psi=
   -\chi \{D r_z +\mu [ r_x \cos(2\psi) +r_y \sin(2\psi) ] \} \varepsilon \cos\psi .
\ear
Equating the coefficients of $\sin\psi$ and $\cos\psi$ in the earlier equation yields
\barl{101} 
   \omega \dot\varepsilon +k\beta^2 \varepsilon_x =-\chi \mu r_y \varepsilon \cos^2\psi ,
\ear
and
\barl{102} 
   2(\dot\phi \omega +k \phi_x)
   -\{ \omega^2 -\Omega^2 -\beta^2 k^2 \}=-\chi [D r_z +\mu r_x \cos(2\psi) ] .
\ear
In order to obtain the dispersion relation $\omega=\omega(k)$, we require the vanishing of the 
expression in the curly brackets in Eq. (\ref{102}), which yields for the wavevector of 
the EM radiation (i.e., the pulse) in the TL the expression
\bel{25} 
   k=\pm \frac{\sqrt{\omega^2-\Omega^2}}{\beta} .
\ee
Thus, the EM wave can propagate through the superconducting quantum metamaterial 
only when its frequency exceeds a critical one,
$\omega_c =\Omega =\sqrt{ (V_{00} +V_{11}) / 2 }$. Finally, Eqs. (\ref{101}) and 
(\ref{102}) are averaged in time over the period $T=2\pi/\omega$ (i.e., the 
fast time-scale) of the phase $\psi (x,t)$. Due to the simple time-dependence of 
$\psi (x,t)$ that has been assumed earlier in the framework of SVEA, that averaging
actually requires the calculation of integrals of the form
\[ \langle {\cal F}(\sin f(\psi),\cos g(\psi) )\rangle =\frac{1}{2\pi}\int^{2\pi}_0 {\cal F}(\sin f(\psi),\cos g(\psi) ) d\psi.\]
This procedure, when it is applied to the two evolution equations Eqs. (\ref{101}) 
and (\ref{102}) gives the truncated equations for slow amplitude and phase
\barl{23}
   \dot\varepsilon +c \varepsilon_x =-\chi \frac{\mu}{2\omega} \varepsilon r_y ,
   \qquad  \dot\phi +c \phi_x=-\chi \frac{D}{\omega}R_z ,
\ear
where $c=\beta^2 k/\omega$ is a critical velocity.

\subsection{Derivation of dynamic equations for the transformed Bloch vector components}
Substituting Eqs. (\ref{21}) and (\ref{22}) into Eqs. (\ref{19}) for the original 
Bloch vector components, we get
\barl{103} 
   (\dot r_x+2\dot{\psi}r_y) \cos(2\psi) +(\dot r_y-2\dot{\psi} r_x)\sin(2\psi)
           =-(\Delta +2D\varepsilon^2 \cos^2\psi) [r_y\cos(2\psi) -r_x\sin(2\psi)]
\ear
\barl{104} 
   (\dot r_y -2\dot{\psi} r_x) \cos(2\psi) -(\dot r_x +2\dot{\psi} r_y) \sin(2\psi)
              =(\Delta +2D\varepsilon^2 \cos^2\psi) [r_x\cos(2\psi) +r_y\sin(2\psi)]
                -2\mu\varepsilon^2 \cos^2\psi r_z
\ear
\bel{105} 
   \dot r_z=2\mu\varepsilon^2 \cos^2\psi [r_y\cos(2\psi) -r_x\sin(2\psi)] .
\ee
In order to derive the dynamic equations for the new Bloch vector components $r_i$,
we first
multiply Eqs. (\ref{103}) and (\ref{104}) by $\cos(2\psi)$ and $\sin(2\psi)$, respectively,
and then subtract one equation from the other. Thus we get 
\bel{106} 
   \dot r_x +2\dot{\psi}r_y
     =-(\Delta +2D\varepsilon^2 \cos^2\psi) r_y +2\mu\varepsilon^2 \cos^2\psi \cos(2\psi) r_z .
\ee
Similarly, by multiplying Eqs. (\ref{103}) and (\ref{104}) by $\sin(2\psi)$ and 
$\cos(2\psi)$, respectively, and by adding them, we get  
\bel{107} 
   \dot r_y -2\dot{\psi} r_x
     =(\Delta +2D\varepsilon^2 \cos^2\psi) r_x -2\mu \varepsilon^2 \cos^2\psi \sin(2\psi) r_z .
\ee
Finally, an average over the phase $\psi$ is performed onto Eqs. (\ref{105})-(\ref{107})
using the (easy to prove) relations $\langle \cos^2\psi \cos(2\psi) \rangle=1/4$ and 
$\langle \cos^2\psi \sin(2\psi) \rangle=0$. That procedure, and using also the relation 
$\dot \psi=\dot \phi-\omega$, yields the truncated Bloch equations
\barl{24}
   \dot r_x=-(\delta +2\dot\phi +D\varepsilon^2)r_y -\frac{\mu}{2} \varepsilon^2 r_z , \qquad
   \dot r_y=+(\delta +2\dot\phi +D\varepsilon^2)r_x , \qquad
   \dot r_z= \frac{\mu}{2} \varepsilon^2  r_y ,
\ear
where $\delta =\Delta -2\omega$.
Note that Eqs. (\ref{24}) possess a dynamic invariant that has a form similar to that 
of the original Bloch equations, Eqs. (\ref{19}), i.e., 
\bel{26} 
   r^2_x+r^2_y + r^2_z=1 .
\ee
The truncated Bloch equations Eqs. (\ref{24}), along with Eqs. (\ref{23}) for the 
slowly varying amplitude and phase of the EM vector potential pulse,
constitute a closed system with five unknown functions which describe the approximate 
dynamics of the superconducting quantum metamaterial that are determined in the
next section.

\section{Exact Integration of the Truncated Equations}
The combination of Eqs. (\ref{23}) and the third equation of Eqs. (\ref{24}) 
provides a relation between the slow amplitude and the phase of the EM vector
potential pulse. The first of Eqs. (\ref{23}) is multiplied by $\varepsilon$ and
written as 
\bel{27}
  {\left[\frac{\partial}{\partial t} +c \frac{\partial}{\partial x} \right]
       \varepsilon^2 (x,t)} =  -\chi \frac{\mu}{\omega} \varepsilon^2 (x,t) r_y ,
\ee
while the derivative of the second of Eqs. (\ref{23}) is taken with respect to 
time and then $\dot{R}_z =\dot{r}_z$ is replaced from Eqs. (\ref{24}). Thus we get
\bel{27b}
   {\left[\frac{\partial}{\partial t} +c \frac{\partial}{\partial x} \right]  
      \dot\phi (x,t)} =-\chi \frac{\mu D}{2 \omega} \varepsilon^2 (x,t) r_y .
\ee
From Eqs. (\ref{27}) and (\ref{27b}), and by taking into account the independence 
of the slow temporal and spatial variables, we get  
\bel{28} 
   2\dot\phi(x,t) = D \varepsilon^2 (x,t) +const. ,
\ee
where the constant of integration can be set equal to zero. Using Eq. (\ref{28}),
the truncated Bloch equations Eqs. (\ref{24}) can be written as 
\barl{29}
  \dot r_x=-(\delta +2 D \varepsilon^2) r_y , \qquad
  \dot r_y=+(\delta +2 D \varepsilon^2) r_x -\frac{\mu}{2} \varepsilon^2  r_z , \qquad
  \dot r_z= \frac{\mu}{2} \varepsilon^2  r_y .
\ear
The latter can be written in a simpler form using the unitary transformation
\barl{108} 
   r_x =S_x \cos\Phi -S_z \sin\Phi , \qquad  r_y =S_y , \qquad 
   r_z =S_z \cos\Phi +S_x \sin\Phi , 
\ear
where $\Phi$ is the constant transformation angle that is going to be determined. 
The truncated Bloch equations for the $r_i$, 
$i=x,y,z$, can be written in terms of the new Bloch vector components $S_i$, 
using a procedure similar to that used in Subsection III.B to obtain Eqs. (\ref{24}).
Sustituting Eqs. (\ref{108}) into Eqs. (\ref{29}), we get 
\bel{110}
   \dot S_x \cos\Phi -\dot S_z \sin\phi=-(\delta +2 D \varepsilon^2) S_y ,
\ee
\bel{109}
   \dot S_y=\left[ (\delta+2D\varepsilon^2) \cos\Phi -\frac{\mu}{2} \varepsilon^2
    \sin\Phi\right] S_x
   -\left[ (\delta + 2 D \varepsilon^2) \sin\Phi +\frac{\mu}{2} \varepsilon^2
    \cos\Phi \right] S_z ,
\ee
\bel{111}
   \dot S_x\sin\Phi +\dot S_z\cos\Phi=-\frac{\mu}{2} \varepsilon^2  S_y .
\ee
Multiplying Eqs. (\ref{110}) and (\ref{111}) by $\cos\Phi$ and $\sin\Phi$, respectively,
and then adding them together, we get 
\bel{200}
   \dot S_x =\left\{ \varepsilon^2 \left[ \frac{1}{2}\mu \sin\Phi -2 D \cos\Phi \right]
                    -\delta \cos\Phi \right\}  S_y .
\ee
Similarly, multiplying Eqs. (\ref{110}) and (\ref{111}) by $\sin\Phi$ and $\cos\Phi$,
respectively, and then subtracting one equation from the other, we get
\bel{201}
   \dot S_z =\left\{ \varepsilon^2 \left[ \frac{1}{2}\mu \cos\Phi +2 D \sin\Phi \right]
                    -\delta \sin\Phi \right\}  S_y .
\ee
The transformation angle $\Phi$ can now be selected so that the resulting equations
are simplified as much as possible. We thus define
\bel{112} 
  \tan \Phi \equiv \gamma =\frac{4 D}{\mu} ,
\ee
so that $\cos\Phi=\pm \sigma$ and $\sin\Phi=\pm \sigma \gamma$ where 
$\sigma =1 / \sqrt{1+\gamma^2}$.
The choice of the sign is irrelevant and here we pick positive sign for both functions.
Using the above value for the transformation angle and the definitions 
$W=\sqrt{(4D)^2 +\mu^2}$ and $\eta= -\delta \mu / W$, Eqs. (\ref{200}), 
(\ref{109}), and (\ref{201}) obtain their final form 
\begin{eqnarray}
\label{97}
   \dot S_x =+\eta S_y , \qquad
   \dot S_y =-\eta S_x +\left[ \eta \gamma -\frac{1}{2} W \varepsilon^2 \right] S_z , \qquad
   \dot S_z =\left[-\eta \gamma +\frac{1}{2} W \varepsilon^2 \right] S_y .
\end{eqnarray}
In order to investigate the possibility for "coherent propagation" of an electromagnetic
potential pulse, we consider resonance conditions. In that case, $\eta=0$, and Eqs. 
(\ref{97}) become
\begin{eqnarray}
\label{202}
   \dot S_x =0 , \qquad
   \dot S_y =-\frac{1}{2} W \varepsilon^2 S_z , \qquad
   \dot S_z =+\frac{1}{2} W \varepsilon^2 S_y .
\end{eqnarray}
Combining the second and third Eq. (\ref{202}) and integrating, we obtain the "resonant"
conservation law $S_y^2 +S_z^2 =const.$. Assuming that all the qubits are in the ground
state at $t=-\infty$, we have the initial conditions $r_x (t=-\infty) =r_y (t=-\infty) =0$ 
and $r_z (t=-\infty) =-1$ which are transformed into 
$S_x (t=-\infty) =-\gamma \sigma$, $S_y (t=-\infty) =0$, and $S_z (t=-\infty) =-\sigma$
through the transformation Eq. (\ref{108}). Applying the initial conditions to the 
resonant conservation law, we get 
\begin{equation}
\label{203}
   S_y^2 +S_z^2 =\sigma^2 .
\end{equation}
In the following, we seek solution of the form $\varepsilon=\varepsilon(\tau=t-x/v)$
and $S_i =S_i (\tau=t-x/v)$, with $i=x,y,z$. By changing the variables in the first of 
Eqs. (\ref{23}), with $r_y$ being replaced by $S_y$, we get  after rearramgement    
\barl{204}
   \frac{\varepsilon_\tau}{\varepsilon} =\chi \frac{\mu}{2\omega} \frac{v}{c-v} S_y .
\ear
Then, combining Eq. (\ref{104}) with the third Eq. (\ref{202}) and integrating, we get
\begin{equation}
\label{205}
   \varepsilon^2 (\tau) =\chi \frac{2\mu}{\omega W}  \frac{v}{c-v} [S_z (\tau) +\sigma] , 
\end{equation}
where the conditions $\varepsilon (-\infty)$ and $S_z (-\infty) =-\sigma$ were used.
The system of Eqs. (\ref{203})-(\ref{205}) for $\varepsilon$, $S_y$, and $S_z$ can be 
integrated exactly; the variables $S_y$ and $S_z$ can be eliminated in favour of 
$\varepsilon$ to give $\varepsilon_\tau =\lambda \varepsilon^2 \sqrt{a +b \varepsilon^2}$,
in which the constants are defined as 
$a=2\sigma/\kappa$, $b=-1/\kappa^2$, $\lambda=\chi \frac{\mu}{2\omega} \frac{v}{c-v}$,
$\kappa=\frac{2\mu}{\omega W}  \frac{v}{c-v}$ to simplify the notation.
The equation for $\varepsilon$ can be readily integrated
\begin{equation}
\label{206}
   \int_{\varepsilon_0}^\varepsilon \frac{d\varepsilon}{\varepsilon^2 \sqrt{a+b \varepsilon^2}}
          =\lambda\int_{\tau_0}^\tau d\tau \Rightarrow
   -\frac{\sqrt{a+b \varepsilon^2}}{a \varepsilon} =\lambda (\tau -\tau_0) ,
\end{equation}
where we have set $\varepsilon_0 \equiv \varepsilon (\tau=\tau_0) =\sqrt{2\sigma \kappa}$ 
to eliminate the boundary term resulting from the integral over $\varepsilon$. Solving 
Eq. (\ref{206}) for $\varepsilon$, we finally get a Lorentzian-like slowly varying 
amplitude 
\begin{equation}
\label{207}
    \varepsilon (\tau) =\frac{\varepsilon_0}{\sqrt{1 +\tau_p^{-2} (\tau -\tau_0)^2  }} , 
\end{equation}
where $\varepsilon_0 =\sqrt{-{a} / {b}} =\sqrt{ 2\sigma \kappa}$ and 
$\tau_p^{-2} =-\frac{a^2 \lambda^2}{b} =\left( \chi \frac{\sigma \mu}{\omega} \right)^2 \left( \frac{v}{c-v} \right)^2$.

\section{Details of the Numerical Simulations and the Role of Decoherence}
The numerical simulations in order to confirm (or not) the existence of the quantum 
coherent effects and the quantum coherence induced in the medium, i.e., the qubit
transmission line, which are predicted through analytical considerations, consist
of integrating Eqs. 3a-3c and 4 of the paper ($n=1,N$) 
\begin{subequations}
\begin{align}
\label{3333a}
  \dot{R}_x(n)&=-\left[ \Delta +8 D \sin^2 \frac{\alpha_n}{2} \right] R_y (n), \\
\label{3333b}
  \dot{R}_y(n)&=\left[ \Delta +8 D \sin^2 \frac{\alpha_n}{2} \right] R_x (n)
                                     -8\mu \sin^2 \frac{\alpha_n}{2} R_z (n) , \\
\label{3333c}
  \dot{R}_z(n)&=+8\mu \sin^2 \frac{\alpha_n}{2} R_y (n) , 
\end{align}
\end{subequations}
and \begin{equation}
\label{3334}
  \ddot\alpha_n 
   +\chi \left[ \Omega^2 +\mu R_x (n) +D R_z (n) \right] \sin \alpha_n =\beta^2 \delta a_n ,
\end{equation}
where $\delta \alpha_n =\alpha_{n-1} -2\alpha_n +\alpha_{n+1}$ and the parameters
$D$, $\Omega$, $\epsilon_p$, $\mu$, $\Delta$, and $\chi$ are defined in the paper
as functions of the interaction matrix elements $V_{ij}$ ($i,j=1,2$), the energy
difference between the ground and the first excited level, $E_1 -E_0$, and the  
discrete vector
potential coupling parameter $\beta$. The earlier system of ordinary differential 
equations is integrated in time with a standard fourth order Runge-Kutta algorithm
with constant time-step $\Delta t$, typically $10^{-3}$. Such small time-steps, 
and even smaller sometimes, are required to conserve up to high accuracy the total 
and partial probabilities
as the compound system of qubits and the electromagnetic vector potential evolve
in time. In all the simulations, periodic boundary conditions are used. 
Due to the particular shape (Lorentzian) of the EM vector potential pulse and the
population inversion pulse in the qubit subsystem, 
very large systems with $N=2^{13} =8192$ and $N=2^{14} =16384$ have been simulated to 
diminish as much as possible the effects from the ends (i.e., the interaction of the 
pulse tail with itself, as it could be in the case of periodic boundary conditions). 
In some cases, it is necessary to simulate even larger systems, with $N=50,000$.
In the figures presented in the paper and here only part of these arrays is shown, 
for clarity.
In order to observe two-photon self-induced transparent (TPSIT) pulses $a_n (t)$ and 
the induced 
quantum coherent population inversion pulses $R_z (n;t)$, the following initial 
coditions are implemented: for the former, the analytically obtained solution 
resulting for the given set of parameters, while for the latter all 
the qubits are set to their ground, $E_0$, state. In terms of the Bloch vector 
variables $R_i$, $i=x,y,z$, that initial condition is specified as:
\begin{equation}
\label{307}
  R_{x,y} (t=0) =0, \qquad R_z(t=0)=-1 , 
\end{equation}
for any $n=1,...,N$. In that case, the above mentioned pulses exist for velocities 
less than the corresponding limiting velocity for TPSIT media, i.e., 
\begin{equation}
\label{3088}
   v < c=\beta^2 \frac{k}{\omega} =2 \beta^2 \frac{k}{\Delta} .
\end{equation}  
The last equality is valid only when {\em the two-photon resonance condition}
$\omega=\Delta/2$
is satisfied. In Eq. (\ref{3338}), $k$ and $\omega$ denote  the wavenumber and frequency
of the carrier wave of the electromagnetic vector potential; while the latter is
determined by the two-photon resonance condition, the former
may vary within an interval which is restricted by the qubit-parameter-dependent 
condition
\begin{equation}
\label{309}
  2 \chi^2 (V_{11}+V_{00}) < (E_1-E_0)^2 ,
\end{equation}
that guarantees the wavenumber $k$ being real.
\begin{figure}[!t]
\includegraphics[angle=0, width=1.0 \linewidth]{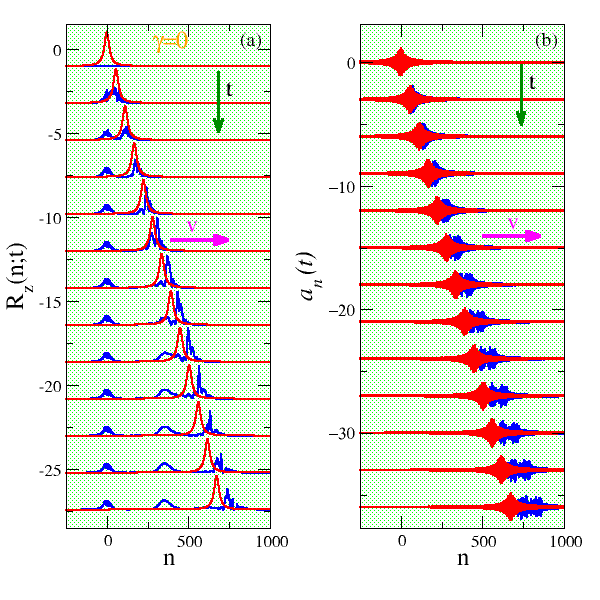}
\caption{
{\bf Numerical validation of the analytical expressions for two-photon self-induced
transparent (TPSIT) propagating pulses in superconducting quantum metamaterials 
(SCQMMs) without decoherence.}
{\bf a,} Snapshots of the population inversion pulse $R_z (n;t)$, excited by the induced
quantum coherence in the qubit subsystem by the electromagnetic vector potential pulse,
in the absence of decoherence ($\gamma=0$); the pulse propagates to the right
(time increases downwards).
{\bf b,} Snapshots for the corresponding evolution of the electromagnetic vector potential
pulse $a_n (t)$, that exhibits significant broadening while its amplitude decreases
as time passes by;
the numerical and analytical pulses are shown in blue and red color, respectively.
Parameter values: 
$\chi=1/4.9$, $\beta=6$, $V_{00}=V_{11}=1$ ($\gamma=0$), $V_{01}=V_{10}=0.7$, 
$E_1 -E_0=3$, and $v/c=0.7$.
}
\label{fig3331}
\end{figure}

\begin{figure}[!t]
\includegraphics[angle=0, width=1.0 \linewidth]{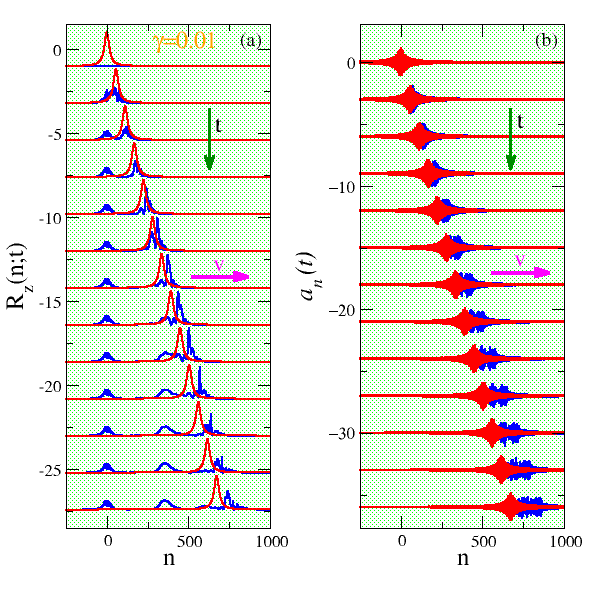}
\caption{
{\bf Numerical validation of the analytical expressions for two-photon self-induced
transparent (TPSIT) propagating pulses in superconducting quantum metamaterials 
(SCQMMs) in the presence of weak decoherence.}
{\bf a,} Snapshots of the population inversion pulse $R_z (n;t)$, excited by the induced
quantum coherence in the qubit subsystem by the electromagnetic vector potential pulse,
in the absence of decoherence ($\gamma=0.01$); the pulse propagates to the right
(time increases downwards).
{\bf b,} Snapshots for the corresponding evolution of the electromagnetic vector potential
pulse $a_n (t)$, that exhibits significant broadening while its amplitude decreases
as time passes by;
the numerical and analytical pulses are shown in blue and red color, respectively.
Parameter values: 
$\chi=1/4.9$, $\beta=6$, $V_{00}=0.998$, $V_{11}=1.002$, 
$V_{01}=V_{10}=0.7$, $E_1 -E_0=3$, and $v/c=0.7$.
}
\label{fig3332}
\end{figure}

\begin{figure}[!t]
\includegraphics[angle=0, width=1.0 \linewidth]{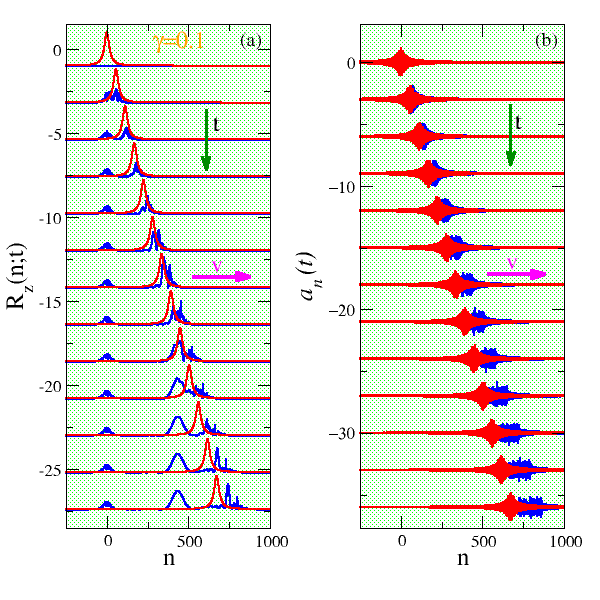}
\caption{
{\bf Numerical validation of the analytical expressions for two-photon self-induced
transparent (TPSIT) propagating pulses in superconducting quantum metamaterials 
(SCQMMs) in the presence of substantial decoherence.}
{\bf a,} Snapshots of the population inversion pulse $R_z (n;t)$, excited by the induced
quantum coherence in the qubit subsystem by the electromagnetic vector potential pulse,
in the absence of decoherence ($\gamma=0.1$); the pulse propagates to the right
(time increases downwards).
{\bf b,} Snapshots for the corresponding evolution of the electromagnetic vector potential
pulse $a_n (t)$, that exhibits significant broadening while its amplitude decreases
as time passes by;
the numerical and analytical pulses are shown in blue and red color, respectively.
Parameter values: 
$\chi=1/4.9$, $\beta=6$, $V_{00}=0.98$, $V_{11}=1.02$, 
$V_{01}=V_{10}=0.7$, $E_1 -E_0=3$, and $v/c=0.7$.
}
\label{fig3333}
\end{figure}

The role of decoherence in TPSIT in superconducting quantum metamaterials like the one
investigated here can be observed in Figures \ref{fig3331}, \ref{fig3332}, and \ref{fig3333}, 
in which the amount of decoherence, quantified by the factor $\gamma$, takes the 
values $0$, $0.01$ and $0.1$, respectively. The decoherence factor $\gamma$ depends
solely on the interaction matrix elements $V_{ij}$; it is given by 
\begin{equation}
\label{310}
  \gamma = 2\frac{V_{11} -V_{00}}{V_{10}} ,
\end{equation}
i.e., it is equal to zero for $V_{00}=V_{11}$ while it assumes non-zero values for 
$V_{11} > V_{00}$. The Figures \ref{fig3331}-\ref{fig3333} show snapshots of $R_z (n;t)$
and $a_n (t)$ at several instants from $t=0$ to $t=168$, which are separated 
by $14$ time units, along with the corresponding analytical solutions.
All these profiles but the first are shifted downwards to avoid overlapping, while only 
part of the array is shown for clarity. Note that time increases downwards.
In Figure \ref{fig3331}, for $\gamma=0$, the $a_n (t)$ pulse is observed to excite a 
population inversion 
pulse $R_z (n;t)$ in the qubit subsystem (Figure \ref{fig3331}a). The amplitude of 
$R_z (n;t)$ gradually increases until it attains its maximum value close to unity,
while at the same time it propagates to the right with velocity $v'$.
In the figure, that occurs for the first time at $t\simeq 70$ time units; subsequently
it evolves in time while it keeps its amplitude almost constant for at least $56$ time
units. After that, its amplitude starts decreasing until it is completely smeared
(not shown). During the time interval in which the amplitude of the $R_z (n;t)$ pulse is 
close to the predicted one (i.e., close to unity), the metamaterial is considered 
to be in the {\em almost quantum coherent} regime. Note that at about $t=84$ a little
bump starts to appear which grows to a little larger ({\em "probability bump"}) which
gets pinned at a particular site at $n \sim 300$. Note that another such bump appears
at $n=0$ due to the initial "shock" of the qubit subsystem because of the sudden onset
of the $a_n (t)$ pulse. A comparison of the numerical $R_z (n;t)$ profiles with the  
analytical ones reveals that the velocity of propagation $v'$, which is the same 
as that of the numerical $a_n (t)$ pulse (Figure \ref{fig3331}b), is slightly larger
than the predicted one $v$ ($v' > v$).
The corresponding profiles for very weak 
decoherence, in which the value of the decoherence factor $\gamma=0.01$, are shown 
in Figure 2. For that weak decoherence, there are only slight differences in 
comparison 
with Figure 1 which actually cannot be observed in the scale of the figures.

In Figure \ref{fig3333}, the decoherence factor has acquired a substantial value of 
$\gamma=0.1$,
so that its effects are now clearly observed in the population inversion pulse 
$R_z (n;t)$, while the vector potential pulse $a_n (t)$ does not seem to be affected.
In Figure \ref{fig3333}a, while the $R_z (n;t)$ is still excited by the vector 
potential pulse $a_n (t)$, it has a very low amplitude compared with that of the 
analytically predicted form. Indeed, that amount of decoherence only slightly
changes the analytical $R_z (n;t)$ pulse and that change is hardly visible in 
the scale of the figure. Note that the speed of the $R_z (n;t)$ is the same as in
the case without decoherence (which is also the same with that of the $a_n (t)$
pulse, in Figures \ref{fig3331}-\ref{fig3333}); even the unwanted "probability bumps" 
appear at about the same locations with almost the same amplitude and shape 
independently of the amount of decoherence.

\begin{figure}[!t]
\includegraphics[angle=0, width=1.0 \linewidth]{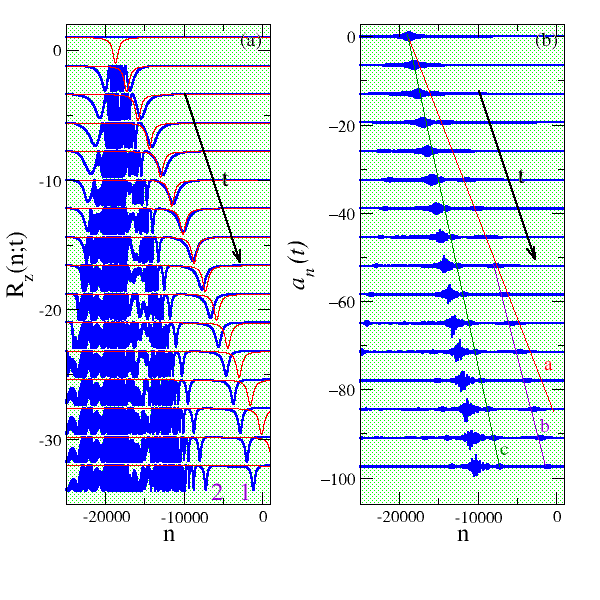}
\vspace{-1cm}
\caption{
{\bf Numerical exploration of the propagation of two-photon superradiant (TPSRD)
pulses in superconducting quantum metamaterials (SCQMMs) for large values of 
$\beta$.}
{\bf a,} Snapshots of the population inversion $R_z (n;t)$ induced in the qubit 
subsystem by the electromagnetic vector potential pulse, $a_n (t)$ (blue), 
in the absence of decoherence ($\gamma=0$). The analytically predicted forms are 
shown in red color.
{\bf b,} Snapshots for the corresponding vector potential pulse $a_n (t)$ at the
same time-instants. 
Parameter values: 
$\chi=1/5$, $\beta=30$, $V_{00}=V_{11}=1$, $V_{01}=V_{10}=0.8$, $E_1 -E_0=3$,
and $v/c=2.5$ 
}
\label{fig4}
\end{figure}

Next, the possibility for the existence of propagating two-photon superradiant 
pulses in SCQMMs for large values of $\beta$ is numerically explored (Figure 4).
In order to obtain that figure, the qubit subsystem is initialized with all the 
qubits in their excited state, while the vector potential pulse assumes its 
analytically predicted form for the selected parameter set. A very large array 
with $N=50,000$ is simulated, and the initial position of the (center of) $a_n (t)$
pulse is at $n=-18,750$. The subsequent evolution
produces the snapshots presented in Figure 4, from $t=0$ up to $t=300$ time units
(time increases downwards).
These snapshots are separated in time by $20$ time units and they are displaced 
vertically to avoid overlapping. The population inversion $R_z (n;t)$
and the electromagnetic vector potential pulse $a_n (t)$ are shown in Figures
\ref{fig4}a and \ref{fig4}b, respectively. In that simulation, the parameter
$\beta$ is much larger than that used in Figure 3b of the paper, i.e., here    
$\beta=30$. The other parameters, except the pulse speed, here $v=2.5 >c$, 
are the same as those used in Figure 3b of the paper. 
The spatio-temporal evolution of 
both pulses reveals a rather complex pattern, that is discussed below.  
In Figure 4b, the $a_n (t)$ pulse breaks into several pulses of different
amplitudes and velocities; the most important are discernible along the green (c)
and the purple (b) solid lines. In the same subfigure, the red (a) 
solid line indicates the analytically predicted pulse speed for two-photon 
superradiance in superconducting quantum metamaterials.
The $a_n (t)$ pulses emerging from the initial one affect the qubit subsystem in a
peculiar and complex way: First of all, a very well-defined $R_z (n;t)$ pulse is
observed which follows the analytical solution (in red) for quite some time
(about $\sim 140$ time units). After that time the pulse (whose last snapshot 
is numbered $1$) slows down while it gets narrower while it is propagating 
together with the $a_n (t)$ pulse along the purple (b) line.
Note that around the initial position of the $a_n (t)$ pulse, the population
inversion $R_z (n;t)$ exhibits a region of strong variability within its extreme 
values $\pm 1$ which expands in the course of time.
At $t \sim 100$ a second $R_z (n;t)$ pulse is observed to emerge from the strong
variability region, which is significantly narrower than the analytical solution.
That second pulse ((whose last snapshot is numbered $2$) propagates slower than 
the first one ($1$), along with the $a_n (t)$ pulse which can be observed along
the green line (c) of Figure \ref{fig4}b.

\end{document}